\documentclass[utf8]{FrontiersinHarvard} 
\usepackage{url,hyperref,lineno,microtype,subcaption}
\usepackage[onehalfspacing]{setspace}
\usepackage{amsmath}

\def\keyFont{\fontsize{8}{11}\helveticabold }
\def\firstAuthorLast{Alford {et~al.}} 
\def\Authors{J. A. J. Alford\,$^{1,*}$, G. A. Younes\,$^{2}$, Z. Wadiasingh\,$^{3,4,5}$, M. Abdelmaguid\,$^{1}$, H. An$^{6}$,  M. Bachetti,$^{7}$, M. Baring$^{8}$, A. Beloborodov$^{9}$, A. Y. Chen$^{10}$, T. Enoto$^{11}$, J. A. Garc\'ia\,$^{12}$, J. D. Gelfand\,$^{1}$, E. V. Gotthelf$^{9}$, A. Harding$^{13}$, C.-P. Hu\,$^{14}$, A.D. Jaodand\,$^{15}$, V. Kaspi\,$^{16}$, C. Kim\,$^{6}$, C. Kouveliotou\,$^{2}$, L. Kuiper\,$^{17}$ K. Mori\,$^{8}$, M. Nynka\,$^{18}$, J. Park\,$^{6}$, D. Stern\,$^{19}$,  J. Valverde\,$^{20}$, D. Walton\,$^{21}$}

\def\Address{\footnotesize $^{1}$NYU Abu Dhabi, PO Box 129188, Abu Dhabi, UAE \\
$^{2}$Department of Physics, The George Washington University, 725 21st St. NW, Washington, DC 20052, USA \\
$^{3}$Astrophysics Science Division, NASA/GSFC, Greenbelt, MD 20771, USA \\
$^{4}$Department of Astronomy, University of Maryland, College Park, Maryland 20742, USA \\
$^{5}$Center for Research and Exploration in Space Science and Technology, NASA/GSFC, Greenbelt, Maryland 20771, USA \\
$^{6}$ Department of Astronomy and Space Science, Chungbuk National University, Cheongju, Republic of Korea\\  
$^{7}$ INAF-Osservatorio Astronomico di Cagliari, Via della Scienza 5, I-09047 Selargius (CA), Italy \\
$^{8}$Department of Physics and Astronomy - MS 108, Rice University, 6100 Main Street, Hous- ton, Texas 77251-1892, USA \\
$^{9}$ Columbia Astrophysics Laboratory, Columbia University, New York, NY 10027, USA \\
$^{10}$ Physics Department and McDonnell Center for the Space Sciences, Washington University in St. Louis; MO, 63130, USA \\
$^{11}$  Extreme Natural Phenomena RIKEN Hakubi Research Team, Cluster for Pioneering Research, RIKEN, 2-1 Hirosawa, Wako, Saitama 351-0198, Japan \\
$^{12}$ X-ray Astrophysics Laboratory, NASA Goddard Space Flight Center, Greenbelt, MD, USA \\
$^{13}$Theoretical Division, Los Alamos National Laboratory, Los Alamos, NM 87545, USA \\
$^{14}$ Department of Physics, National Changhua University of Education, Changhua 50007, Taiwan \\
$^{15}$Division of Physics, Mathematics, and Astronomy, California Institute of Technology, Pasadena, California, USA \\
$^{16}$Department of Physics, McGill University, 3600 rue University, Montreal, QC H3A 2T8, Canada \\
$^{17}$SRON-Netherlands Institute for Space Research, Niels Bohrweg 4, 2333 CA, Leiden, Netherlands \\
$^{18}$Kavli Institute For Astrophysics and Space Research, Massachusetts Institute of Technology, Cambridge, MA 02139, USA \\
$^{19}$Jet Propulsion Laboratory, California Institute of Technology, 4800 Oak Grove Dr., Pasadena, CA 91109, USA \\
$^{20}$Department of Physics and Center for Space Sciences and Technology, University of Maryland Baltimore County, Baltimore, MD 21250, USA \\
$^{21}$Centre for Astrophysics Research, University of Hertfordshire, College Lane, Hatfield, AL10 9AB, UK}

\def\corrAuthor{J. A. J. Alford}

\def\corrEmail{alford@nyu.edu}

\begin{document}
\onecolumn

\title[HEX-P Magnetars and Isolated Neutron Stars]{The High Energy X-ray Probe (HEX-P): Magnetars and Other Isolated Neutron Stars}

\author[\firstAuthorLast ]{\Authors} 
\address{} 
\correspondance{} 

\extraAuth{}

\maketitle
\begin{abstract}

The hard X-ray emission from magnetars and other isolated neutron stars remains under-explored.
An instrument with higher sensitivity to hard X-rays is critical to understanding the physics of neutron star magnetospheres and also the relationship between magnetars and Fast Radio Bursts (FRBs).
High sensitivity to hard X-rays is required to determine the number of magnetars with hard X-ray tails, and to track transient non-thermal emission from these sources for years post-outburst.
This sensitivity would also enable previously impossible studies of the faint non-thermal emission from middle-aged rotation-powered pulsars (RPPs), and detailed phase-resolved spectroscopic studies of younger, bright RPPs.
The High Energy X-ray Probe (HEX-P) is a probe-class mission concept that will combine high spatial resolution X-ray imaging ($<5$ arcsec half-power diameter (HPD) at 0.2--25 keV) and broad spectral coverage (0.2--80 keV) with a sensitivity superior to current facilities (including XMM-Newton and NuSTAR).
HEX-P has the required timing resolution to perform follow-up observations of sources identified by other facilities and positively identify candidate pulsating neutron stars.
Here we discuss how HEX-P is ideally suited to address important questions about the physics of magnetars and other isolated neutron stars.

\section{}

\tiny
 \keyFont{ \section{Keywords:} X-ray sources, HEX-P, pulsars, magnetars, neutron stars, spectra} 
\end{abstract}
\section{Introduction}

\begin{figure}[h!]
\begin{center}
\includegraphics[width=0.5\textwidth]{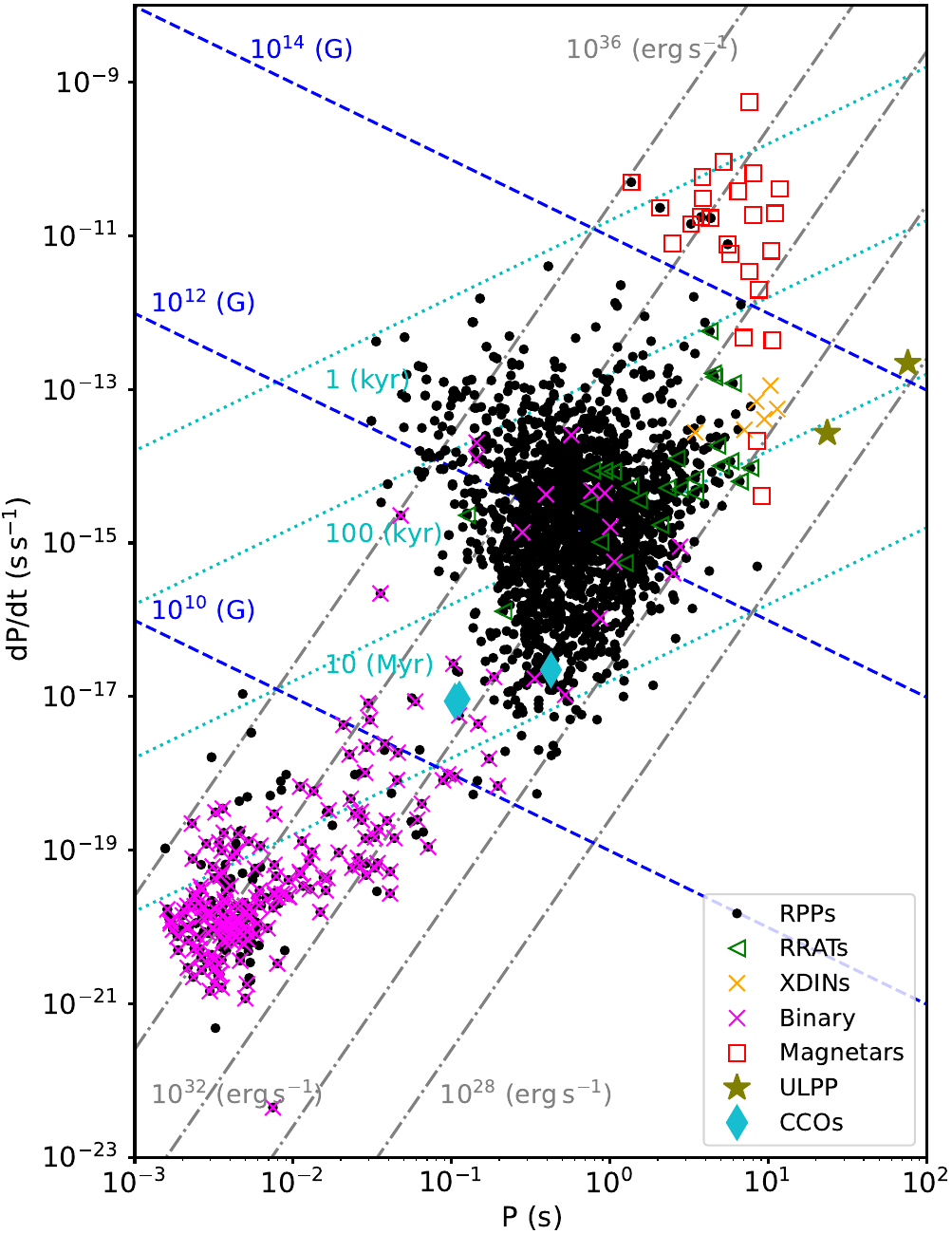}
\end{center}
\caption{Pulsar period period-derivative diagram highlighting the range of neutron star classes. Data are from the ATNF pulsar catalog \citep{Manchester2005}.}\label{fig:ppdot}
\end{figure}

Massive stars ($\gtrsim8$~$M_{\odot}$) end their lives in spectacular supernovae, leaving behind either a neutron star (NS) or a black hole (BH) \citep{Heger2003}.
NSs are very compact (having masses of $\approx1.4$~$M_{\odot}$ and radii of $\approx10$ km), and isolated NSs typically have strong magnetic fields ($B\gtrsim10^{12}$ G, though see the discussion of central compact objects (CCOs) below).
The telltale sign for the existence of an NS is the detection of their spin ephemerides, most notably their spin periods $P$ and, if multi-epoch observations exist, their period derivatives $\dot{P}$. These two observational properties can be used to estimate three key fundamental physical scales under the assumption of a rotating dipole magnet in vacuum (Figure \ref{fig:ppdot}): (1) rotational energy loss $|\dot{E}| \propto \dot{P}/P^{3}$, (2) characteristic spin-down age $\tau_{\rm char} = P/2\dot{P}$, and (3) surface dipolar magnetic field strength $B_{\rm dip} \propto \sqrt{P\dot{P}}$.

Since their discovery, a number of unique classes of isolated NSs (INSs) have been identified through differences in their broadband emission characteristics, location/environment, and timing behavior. The most common of the INS population are the rotation-powered pulsars (RPPs), aptly named given that their large spin-down power $\dot{E}$ far exceeds their total radiative luminosity. Then, there are the magnetars, which possess periods typically in the range of about $1$-$15$ seconds and large spin-down rates ($\dot{P} \sim 10^{-11}$ s s$^{-1}$) , thus occupying a unique space in the $P$-$\dot{P}$ diagram (Figure \ref{fig:ppdot}, red squares). Assuming magnetic dipole braking, the magnetar timing properties translate to dipole field strengths of the order of $10^{14}$~G (two orders of magnitude larger than RPPs, \citealt{kouveliotou98Natur,kouveliotou99ApJ,kaspi03ApJ}), average spin-down ages of $10^4$~years (confirmed through the association of a few with young, X-ray bright supernova remnants (SNRs), see, e.g., \citealt{Vasisht1997}), and low spin-down power $|\dot{E}| \sim 10^{32}$~erg~s$^{-1}$ \citep{Olausen2014}. Despite the latter, magnetars are observed as hot thermal emitters with surface thermal temperatures $kT\approx0.5$~keV (factors of $\sim$2 larger than young RPPs; see, e.g., Figure~14 of \citealt{Olausen2014}), and X-ray luminosities $\gtrsim10^{33}$~erg~s$^{-1}$. 
Hence, magnetar radiative power is attributed to the decay of their extreme external and internal $B$-fields \citep{Thompson1995,kaspi17ARAA}.

Then there are the dozen known central compact objects (CCOs), which are perhaps the least understood class of INSs \citep{DeLuca2017}. The three CCOs with measured spin ephemerides are indicated with blue diamonds in Figure \ref{fig:ppdot}. 
These point-like X-ray sources are found in young X-ray bright SNRs, and lack both emission at other wavelengths and any associated pulsar wind nebula. While there are relatively few identified CCOs, their locations in young SNRs suggests that they may represent a significant fraction of NS births. 
CCOs are also known as `anti-magnetars', since some of them have the smallest spin-down-measured dipole magnetic field strengths among all known young NSs \citep{Gotthelf2013}, at odds with their relatively bright thermal X-ray emission. 

X-ray dim isolated NSs (XDINSs, yellow crosses in Figure \ref{fig:ppdot}) constitute a population with observational characteristics that do not quite fit the above classes. They are nearby (within a few hundred parsecs, \citealt{Kaplan2008}), radio-quiet \citep[but see][]{Rigoselli19AA}, yet thermally-emitting with surface temperatures in the range of $45$-$110$~eV and luminosities $L_{\rm X} < 10^{32}$~erg~s$^{-1}$. 
This emission is pulsed at a spin-period of a few seconds, while slowing down with a rate of $\dot{P}\approx10^{-14}$~s~s$^{-1}$, implying $B_{\rm dip}\sim10^{13}$~G, $\tau\gtrsim1$~Myr, and an average $\dot{E}\sim5 \times 10^{30}$~erg~s$^{-1}$. 
Given their small $|\dot{E}|$ compared to their surface thermal emission, XDINS are thought to be powered by their relatively large $B$-fields, akin to their younger counterparts, magnetars.

Recently, a new class of long-period radio-emitting NSs have been discovered with spin periods of tens of seconds and relatively large spin-down rates (\citealt{tan18ApJ,caleb22NatAs}; green stars in Figure~\ref{fig:ppdot}). Their dipole-inferred fields are comparable to those of magnetars and XDINs. They are also relatively nearby with distances of the order of 1~kpc, yet deep X-ray observations have not detected their high energy counterpart with upper-limits comparable to those of XDINs \citep{rea22ApJ,beniamini23MNRAS}. These limits are consistent with the old magnetar interpretation, though how they maintain dipolar-fields of $10^{14}$~G is less clear \citep{caleb22NatAs,beniamini23MNRAS}.

Our understanding of the young INS population has been predominantly shaped by radio and X-ray observations. Radio surveys of large areas of the Galaxy provide understanding of the birth rate and properties of NSs \citep[e.g.,][]{arzoumanian02ApJ,faucher06ApJ}, while X-ray observations established several of the above sub-classes, and highlighted the connections among them; e.g.,  RPPs and CCOs showing magnetar-like bursting and outburst abilities \citep{gavriil08Sci,Archibald16ApJ, rea16ApJ}, the existence of low $B$-field magnetars \citep{rea2010Sci}, and bona-fide magnetars with RPP characteristics \citep{camilo06Natur,camilo07ApJ,younes2016ApJ}. All of these X-ray discoveries have led to several attempts at unifying these sub-classes \citep{kaspi10ApJ}, mainly through magneto-thermal evolution models \citep{vigano13MNRAS,gourgouliatos2016PNAS}. Lastly, in a breakthrough discovery, the long-held view that Fast Radio Bursts (FRBs) might be powered by magnetar activity was confirmed after the detection of an FRB \citep{chime2020:1935, Bochenek20:1935} simultaneous with a hard X-ray burst \citep[e.g.,][]{mereghetti20ApJ,li21NatAs} from a Galactic magnetar in outburst.

In this paper, we present the case for the probe-class mission HEX-P as the next generation X-ray satellite to build upon the legacy of all past and current X-ray satellites, and open up avenues for new discoveries in the timely field of high energy studies of young INSs.  A companion paper, Ludlam et al. (in prep.), discusses the power of HEX-P to study neutron stars in low- and high-mass binary systems.

\section{MISSION DESIGN}

The High-Energy X-ray Probe (HEX-P; Madsen et al. 2023) is a probe-class mission concept that offers sensitive broad-band coverage ($0.2-80$\,keV) of the X-ray spectrum with exceptional spectral, timing and angular capabilities. It features two high-energy telescopes (HETs) that focus hard X-rays and one low-energy telescope (LET) that focuses lower-energy X-rays.

The LET consists of a segmented mirror assembly coated with Ir on monocrystalline silicon that achieves a half power diameter of 3.5”, and a low-energy DEPFET detector, of the same type as the Wide Field Imager \citep[WFI;][]{Meidinger2020} onboard Athena \citep{Nandra2013}. 
It has 512 $\times$ 512 pixels that cover a field of view of 11.3' $\times$ 11.3' (1.32$^{"}$ / pixel). It has an effective passband of $0.2-25$\,keV, and a full frame readout time of 2\,ms, which can be operated in a 128 and 64 channel window mode for higher count-rates to mitigate pile-up and faster readout. 
Pile-up effects remain below an acceptable limit of $\sim 1\%$ for fluxes up to $\sim 100$\,mCrab in the smallest window configuration.
Excising the core of the point spread function (PSF), a common practice in X-ray astronomy, will allow for observations of brighter sources, with a typical loss of up to $\sim 60\%$ of the total photon counts.

The HET consists of two co-aligned telescopes and detector modules. The optics are made of Ni-electroformed full shell mirror substrates, leveraging the heritage of XMM-Newton \citep{Jansen2001}, and coated with Pt/C and W/Si multilayers for an effective passband of $2-80$\,keV. 
The high-energy detectors are of the same type as flown on NuSTAR \citep{Harrison2013}, and they consist of 16 CZT sensors per focal plane, tiled 4 $\times$ 4, for a total of 128 $\times$ 128 pixel spanning a field of view of 13.4' $\times$ 13.4'.

The broad X-ray passband and superior sensitivity will provide a unique opportunity to study INSs across a wide range of energies, luminosity, and dynamical regimes.

\section{SIMULATIONS}
\label{hexpSim}

All the simulations presented here were produced with a set of response files that represent the observatory performance based on current best estimates as of Spring 2023 (see Madsen+23). The effective area is derived from a ray-trace of the mirror design including obscuration by all known structures. The detector responses are based on simulations performed by the respective hardware groups, with an optical blocking filter for the LET and a Be window and thermal insulation for the HET. The LET background was derived from a GEANT4 simulation \citep{Eraerds21} of the WFI instrument, and the HET background was derived from a GEANT4 simulation of the NuSTAR instrument, with both simulations adopting a Lagrange point L1 orbit for HEX-P. 

We utilize Xspec version 12.13.0c to simulate point source spectra, implementing HEX-P LET and HET response matrices, ancillary, and background files version v07 \citep{Arnaud1996}. 
All simulated spectra are binned to have 5 counts per energy channel. For spectral fitting, we utilize the Cash-statistic (C-stat in Xspec, \citealt{cash79ApJ}), to derive the best fit model parameters and corresponding uncertainties. 
To assess the goodness of fit, we utilize the \texttt{goodness} command which simulates 1000 spectral realizations from a given model and compares their fit statistic to that of the data; if the data is drawn from the model, or, in other words, the model is a good fit to the data, the fit statistic should lie around the $50\%$ mark.

\section{Magnetars}

\subsection{Persistent broadband X-ray emission}

Magnetar persistent emission consists of two components, a thermal emission likely emanating from surface hot-spots with temperatures of $\sim$ 0.4 -- 0.5~keV, and a non-thermal component with photon index $\Gamma=0.0-1.5$, i.e., rising in $\nu F_{\rm \nu}$, known as the hard X-ray tail. The latter likely originates from inverse Compton scattering of soft photons by relativistic electrons in non-potential magnetospheric loops energized by twists and currents, tied to footpoints whose evolution is driven internally by the crust of the magnetar. Due to the large magnetic fields, the scattering in the magnetosphere is resonant at the electron cyclotron frequency, which is efficient at boosting the photon energies by orders of magnitude relative to the non-resonant case, a process known as Resonant Inverse Compton Scattering (RICS).

\begin{figure*}[t!]
\begin{center}
\includegraphics[width=\linewidth]{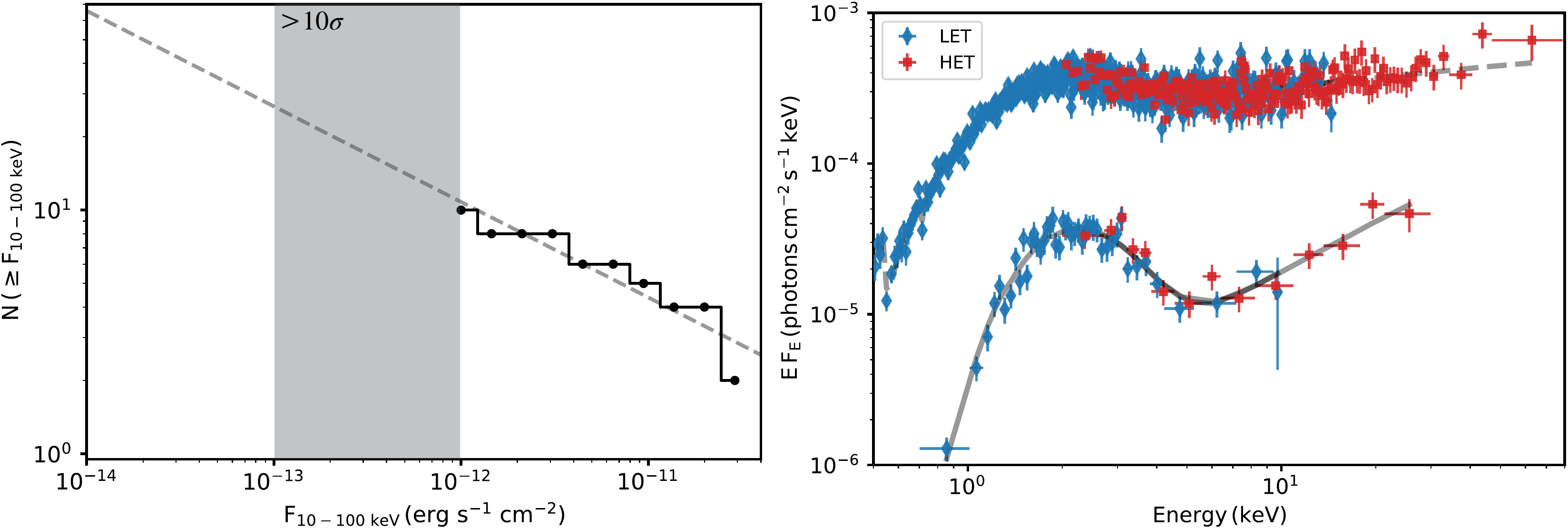}
\end{center}
\caption{{\sl Left panel.} Magnetar $\log~N-\log~S$ distribution in the hard (10 -- 100~kev) X-ray band. NuSTAR doubled the number of magnetar detected at $>10$~keV, which currently stands at 10. The faintest hard X-ray tail known has a flux of the order $10^{-12}$~erg~$s^{-1}$~cm$^{-2}$ \citep[SGR 0526$-$66,][]{park2020ApJ}. HEX-P, with its superior sensitivity, will triple the current number of hard X-ray tails detected from magnetars (gray shaded area). {\sl Right panel.} 100~ks HEX-P simulation of the SGR 0526$-$66 broad-band spectrum and of a generic magnetar with a hard X-ray flux in the 10-100~keV band an order of magnitude weaker, i.e., $10^{-13}$~erg~s$^{-1}$~cm$^{-2}$ (see main text for details).}\label{fig:perEmis}
\end{figure*}

Hard X-ray tails were first discovered in 2004 with INTEGRAL and RXTE \citep{kuiper04ApJ}. Prior to the NuSTAR launch, only a handful of magnetars were detected at energies $>10$~keV \citep{kuiper06ApJ,denhartog08AA,denhartog08AA:1708, enoto10ApJ}. NuSTAR, with its superior sensitivity at hard X-rays, has doubled the pool of detected sources, which is currently standing at 10 \citep[e.g.,][]{enoto17ApJS}.  The faintest known magnetar hard X-ray tail has a 10-100 keV flux of $\sim 10^{-12}$~erg~s$^{-1}$~cm$^{-2}$; an order of magnitude fainter than detections by NuSTAR predecessors. HEX-P, owing to its superior sensitivity, simultaneous broad-band coverage, and more efficient orbit, will detect magnetar hard X-ray tails an order of magnitude fainter again, and characterize their spectral curvature (i.e. a possible energy dependence of the photon index $\Gamma$). This is demonstrated through a 100~ks HEX-P simulation of a typical magnetar spectrum, i.e., $kT=0.45$~keV, $\Gamma=1.0$, and $N_{\rm H}=10^{22}$~cm$^{-2}$, with absorption-corrected fluxes in the 1--10 keV and 10--100 keV band of $5.0\times10^{-14}$~erg~s$^{-1}$~cm$^{-2}$ and $10^{-13}$~erg~s$^{-1}$~cm$^{-2}$, respectively (Figure~\ref{fig:perEmis}). The former flux level represents the faint end of the soft X-ray flux level from the currently known magnetar population while the latter is an order of magnitude fainter than the faintest magnetar hard X-ray tail known \citep[SGR~0526$-$66;][]{park2020ApJ}. The hard tail is detected up to 35 keV with a count rate of $(2.3\pm0.2)\times10^{-3}$~counts~s$^{-1}$ (i.e., a $10\sigma$ detection significance), and the spectral curvature of the hard tail is constrained to $30\%$ ($\Gamma=1.0\pm0.3$). For comparison, we also perform a 100~ks simulated observation of the broad-band X-ray spectrum of the magnetar SGR 0526$-$66 in the Large Magellanic Cloud based on results from Chandra and NuSTAR \citep{park2020ApJ}. HEX-P observations will produce a 10-80~keV count rate of $(44.3\pm0.2)\times10^{-3}$~counts~s$^{-1}$ and a 0.5-10~keV rate of  $(11.6\pm0.1)\times10^{-2}$~counts~s$^{-1}$, providing excellent quality data for a detailed spectral and temporal analysis.

According to the current magnetar $\log{N}$-$\log{S}$ distribution, where $S$ is the 10-100 keV flux, the HEX-P sensitivity limit will enable hard X-ray tail detection and characterization in about 30 magnetars, tripling the current number. Consequently, this will permit a far more comprehensive population-wide correlation analysis between the soft and hard X-ray properties \citep{marsden01ApJ, kaspi10ApJ, enoto10ApJ, enoto17ApJS, Seo2023JKAS}, in turn informing on the evolution of internal and external $B$-field and globally or locally twisted magnetospheres \citep{beloborodov09ApJ,parfrey13ApJ,vigano13MNRAS,CB-2017-ApJ}.

\begin{figure*}[t!]
\begin{center}
\includegraphics[width=\linewidth]{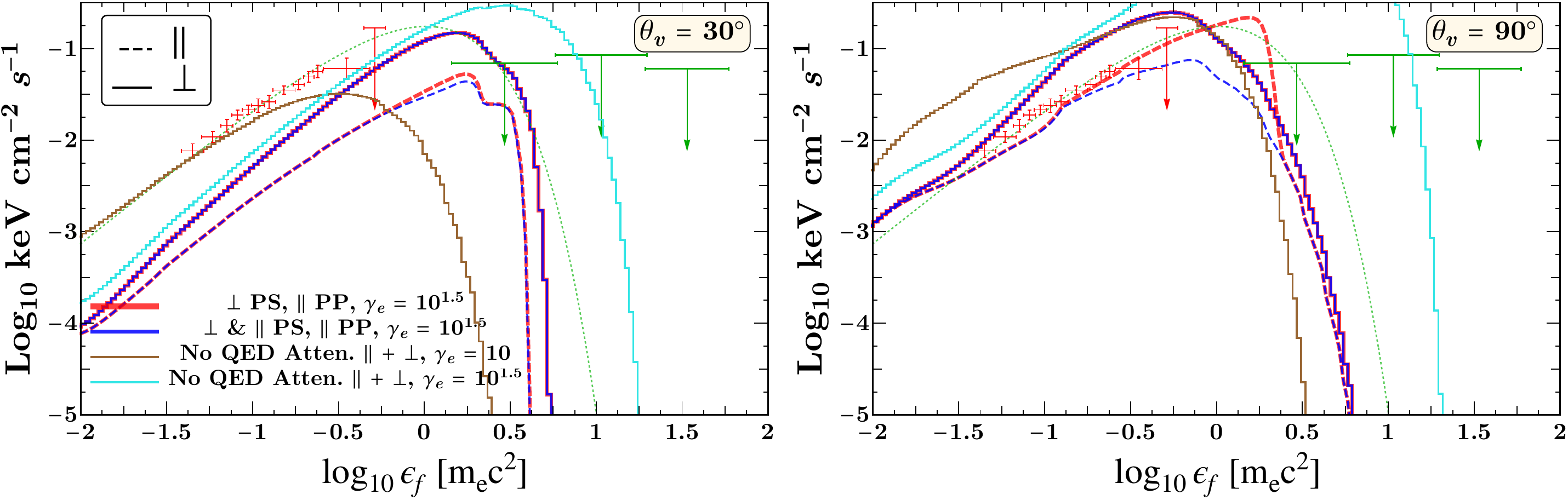}
\end{center}
\caption{Phase-resolved RICS spectra overlaid on INTEGRAL data for 4U 0142+61 (orange data points, \citealt{kuiper06ApJ}), along with a power-law with exponential cutoff
at 350 keV (dotted green). The model emission is computed for surface photons of temperature $5\times10^6$~K
scattered by $\gamma_e=10$-$10^{1.5}$ electrons uniformly populating field bundle from magnetic footpoint colatitudes 12-45$^\circ$. The assumed dipole field strength is $4\times10^{14}$~G. Brown and dark blue lines show total intensities, while red and light blue curves show the case with QED attenuation. {\sl Left panel.} Observer angle to magnetic axis $\theta_v=30^\circ$. {\sl Right panel.}  $\theta_v=90^\circ$. The RICS emission is predicted to be highly phase dependent, as shown here. Adapted from \citet{Wadiasingh19BAAS}.}
\label{fig:rics}
\end{figure*}

For the brightest magnetars, HEX-P will also provide the most detailed look at the 0.3-80~keV spectro-temporal properties that are crucial to guide the theoretical development of radiation transport in the high B-field ($>10^{14}$~G) regime, inaccessible to terrestrial laboratories, such as photon splitting. 
Early predictions for the phase-resolved spectra and energy dependent profiles in the 10--80 keV band are presented in \citet[][also see: \citealt{2007Ap&SS.308..109B,2007ApJ...660..615F,Beloborodov2013ApJ, Caiazzo2022MNRAS, Taverna2020MNRAS}]{Wadiasingh2018ApJ}, and more sophisticated models are in development \citep[e.g.,][]{Wadiasingh2019BAAS,2022HEAD...1911047W}. 
Yet, the current data quality above 10~keV for even the brightest magnetars  (i.e., 4U 0142+61 and 1RXS J170849.0$-$400910), is inadequate  for the detailed phase-resolved spectroscopy required to confront these models. 
HEX-P, providing far superior broad-band data for the brightest magnetars, will allow us to answer fundamental questions which currently remain open: (1) Where are the locales of particle acceleration within magnetar magnetospheres? (2) What are the Lorentz factors of the energetic particles? (3) How do the physical properties governing the hard X-ray tails evolve with age and field strength? (4) Is the hard X-ray emission for persistently emitting magnetars dominated by the dipolar field, or do higher order, crustal fields dominate?

RICS emission is highly anisotropic and sensitive to where cyclotron resonance in the magnetosphere is sampled by an observer. Moreover, beginning around 30 keV, photon splitting can begin to impact spectra depending in the viewing angle or, equivalently, pulse phase \citep{2019MNRAS.486.3327H,2022HEAD...1911047W,2022ApJ...940...91H}. As such, RICS radiation models may obtain a variety of phase-dependent spectral energy distributions (or equivalently, energy dependent pulse profiles) depending on viewing geometry and zones of activation of relativistic particle populations (Figure~\ref{fig:rics}). Detailed fitting of phase-resolved spectra to models of RICS emission is sensitive to the activated zones and observer viewing geometry, providing answers to the open questions laid out above. Furthermore, a comparison of viewing and field geometries will test related constraints obtained for soft thermal emission hot spot modeling from IXPE observations of bright magnetars \citep{Taverna2022Sci,Zane2023ApJ}. 

RICS constraints on the relativistic electron population, along with 0.2--80~keV broadband spectroscopy provided by HEX-P will also test if return currents and particle bombardment play a significant role in heating surface layers of magnetars.
Moreover, HEX-P will provide a detailed population-level phase-resolved spectral survey of magnetars.
This will inform evolutionary traits in the RICS parameters with age, and determine if beaming is compatible with the lack of observed hard X-ray emission in some moderately bright X-ray magnetars.
We note that fitting phase-resolved spectra with RICS models has not yet been attempted due to the paucity of pulsed counts at high energies ($>10$~keV).
For instance, the brightest magnetar at hard X-rays, 1E~1841$-$045, has a {\sl NuSTAR} count rate in the 10--79 keV band of $0.16$~counts~s$^{-1}$, which, for the 350~ks existing observation \citep{An2015ApJ}, results in 56,000 phase-averaged counts, and 11,200 pulsed counts.
For a modest phase-resolved spectroscopic analysis with 10 phase bins, the 1100 counts in each bin were able to constrain the hard X-ray photon index to about $\approx20\%$ \citep{An2015ApJ}.
In contrast, the HEX-P count rate for 1E~1841$-$045 in the same energy range is predicted to be 0.62~counts~s$^{-1}$, which, for the same considerations above, would result in $\sim$ 4350 counts per phase bin, allowing us to search for phase variability in the hard X-ray tail down to the $\lesssim5$ percent level, notably aided by the LET instrument which will constrain the soft thermal part of the spectrum.
Furthermore, observing from L1 rather than from low-Earth orbit like NuSTAR, HEX-P has nearly twice the observing efficiency of NuSTAR. 

Lastly, we note that magnetars have attracted interest from the dark matter community as testbeds for certain axion-like particle (ALP) models.
ALPs produced in the magnetar core are predicted to convert into photons in the magnetosphere.
ALP models generally predict an opposite hard X-ray phase dependence to RICS, thereby enabling HEX-P to provide important constraints to the dark matter community \citep[e.g.,][]{Maruyama2018PhLB,Fortin2021JCAP}.

\subsection{Transient behavior - crustal and magnetospheric dynamics}
\label{transMag}

\begin{figure*}[h!]
\begin{center}
\includegraphics[width=1.0\linewidth]{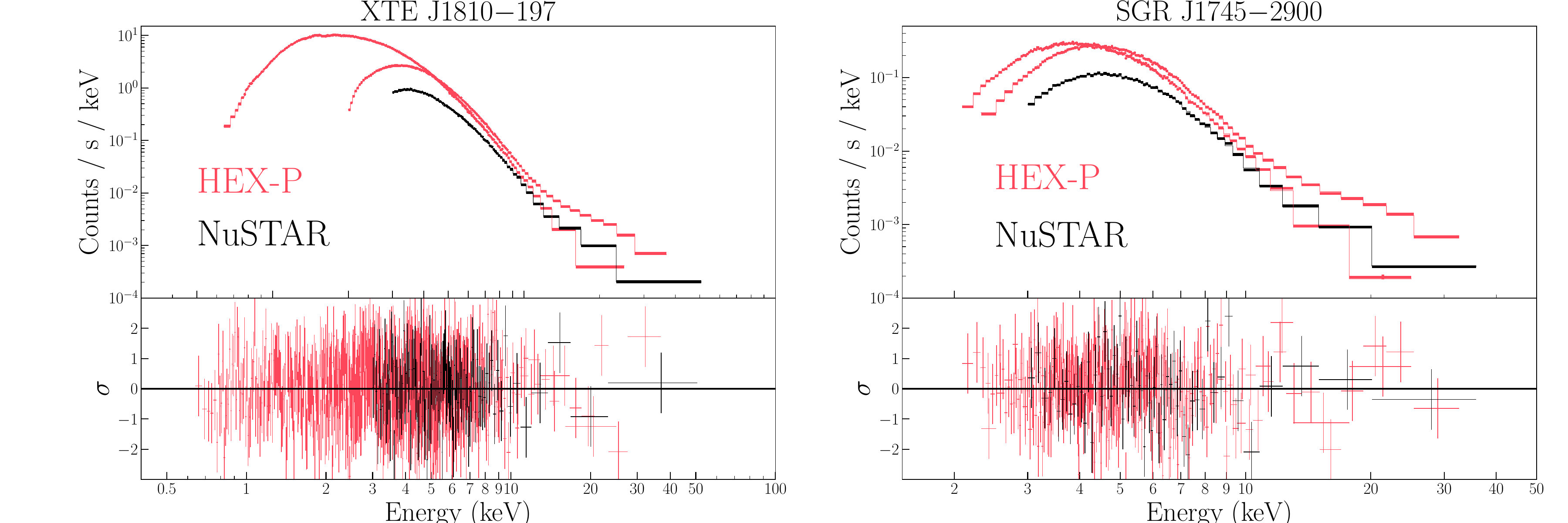}
\end{center}
\caption{Simulated HEX-P broadband X-ray spectra of two transient magnetars near the peak of their outbursts: XTE J1810$-$197 (left) and SGR J1745$-$2900 (right) based on data from their 2018 and 2013 outbursts, respectively.
LET and HET data are shown, with the LET data distinguished by its  higher flux at low energies.
Note that SGR J1745$-$2900 is highly absorbed due to its location in the Galactic center region.
}\label{fig:transient_magnetar_2}
\end{figure*}

\begin{figure*}[h!]
\begin{center}
\includegraphics[width=.9\linewidth]{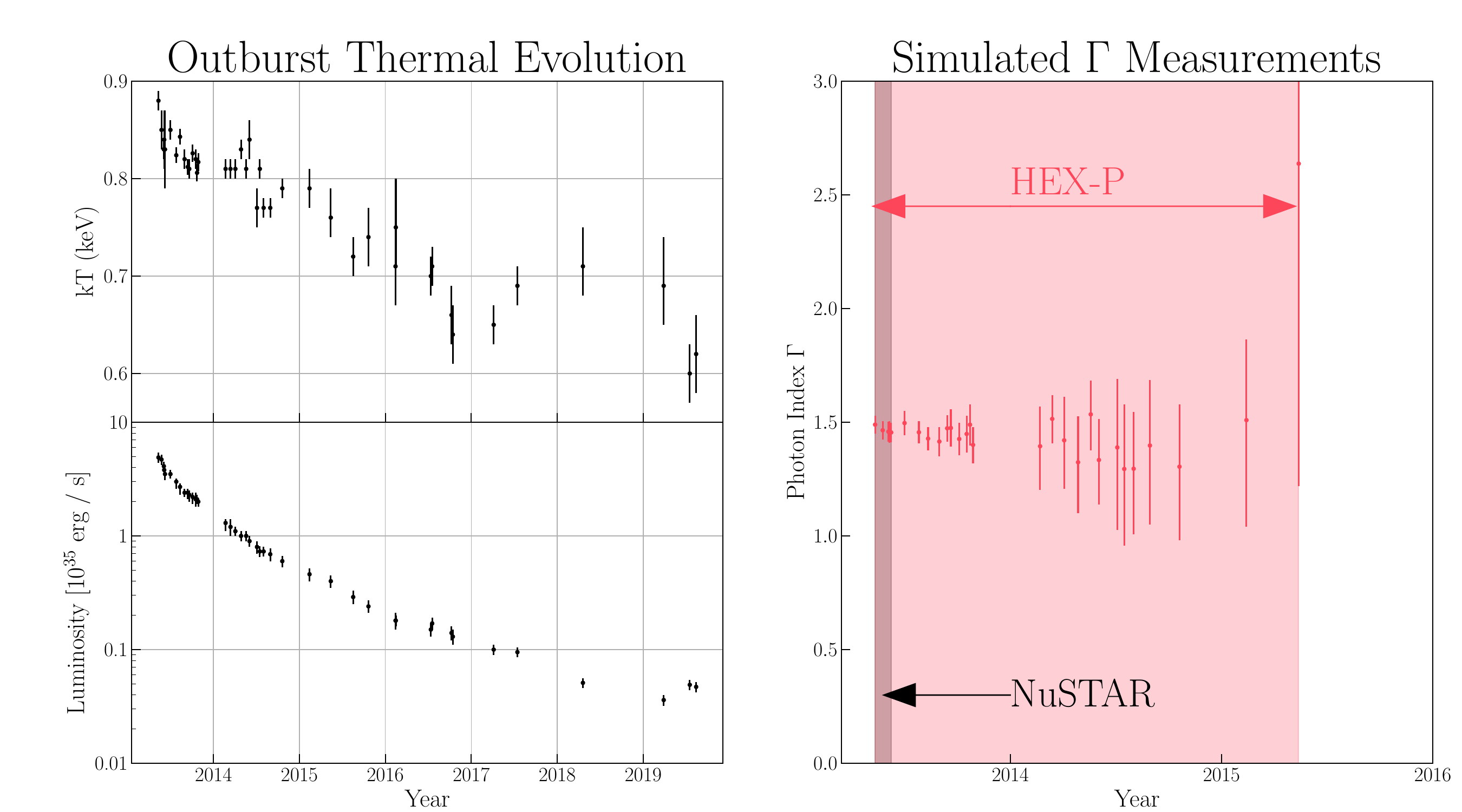}
\end{center}
\caption{Simulated HEX-P measurements of the photon index $\Gamma$ following the 2013 outburst of SGR 1745$-$2900 based on NuSTAR and Chandra observations. The upper panel of the left figure shows the evolution of the blackbody temperature, where the evolution of the 0.3$-$10.0 keV X-ray luminosity is shown in the lower panel of the left figure. The right figure shows the evolution of $\Gamma$ obtained with NuSTAR and HEX-P. HEX-P's improved sensitivity will enable monitoring of these outbursts significantly longer than NuSTAR.
}\label{fig:transient_magnetar_1}
\end{figure*}

In addition to the persistent magnetars, with X-ray luminosities reaching $\sim 10^{35}$ erg s$^{-1}$, there are also `transient' magnetars, which only reach $\sim 10^{35}$ erg s$^{-1}$ during outbursts, when their X-ray fluxes increase by over two orders of magnitude \citep{coti18MNRAS}. Transient magnetar outbursts decay on exponential timescales ranging from months to years \citep{Rea2009,Kaspi2014,coti18MNRAS}. The outburst spectra feature both surface thermal emission and non-thermal magnetospheric emission.

During the decay phase, thermal hot spots are observed to cool and shrink in size.
This behaviour is consistent with currents circulating through a twisted magnetosphere, depositing heat at the surface footpoints of magnetic current loops \cite{beloborodov09ApJ}. Alternatively, this behavior might be due to internal heating of the crust through magnetic stresses associated with evolving toroidal fields \citep{Lander2015MNRAS,lander19MNRAS,2023ApJ...947L..16L}. While the decay of the thermal emission has been well observationally constrained, the hard, non-thermal X-ray emission fades beyond current detection limits fairly quickly (over the course of weeks). It is therefore unknown whether the non-thermal emission decays in tandem with the soft X-ray emission, as expected in the case of surface bombardment by accelerated particles in the magnetosphere, or whether the two evolve independently. This might be the case if the surface heating is indeed induced internally \citep[e.g.,][]{Kouveliotou2004ApJ,Pons2012ApJ,Deibel2017ApJ}, independent of the external magnetospheric emission. This is a major open question in transient magnetars given its potential to investigate crustal micro- and macro-physical properties, which are poorly known, and are highly relevant to the NS equation of state. HEX-P's sensitivity and broadband X-ray coverage are uniquely capable of addressing this fundamental open question.

The left panel of Figure \ref{fig:transient_magnetar_2} shows simulated HEX-P and NuSTAR spectra based on the X-ray spectrum of the transient magnetar XTE J1810-197 during its late 2018 outburst \citep{Gotthelf2019}. This highlights the essential role that the LET plays in constraining the soft $<2$~keV spectrum, especially in the case of less absorbed sources such as for XTE J1810-197.

The right panel of Figure \ref{fig:transient_magnetar_2} shows HEX-P simulations of the hard X-ray spectrum of decaying flux from SGR J1745$-$2900 following an outburst, based on NuSTAR and Chandra observations of the 2013 outburst \citep{Kaspi2014}.
The soft thermal emission follows the multi-year evolution observed by Chandra \citep{Rea2020}.  The luminosity of the non-thermal emission is set equal to what was observed by NuSTAR at the beginning of the outburst, and is assumed to decay with the thermal luminosity. While NuSTAR tracked the SGR J1745$-$2900 non-thermal emission for only four months post-outburst, HEX-P will be capable of tracking a similar magnetar outburst for two years (Figure~\ref{fig:transient_magnetar_1}). This will provide data critical to our understanding of NS crustal and magnetospheric dynamics, unique observational characteristics of magnetars following an outburst.

\subsection{Fast Radio Bursts}

Fast Radio Bursts (FRBs) are millisecond, bright radio bursts (fluence $\sim$ few Jy ms) observed over a broad range of frequencies, from $\sim$120 MHz to a few GHz. They were first reported in 2007 \citep{lorimer07Sci}, and since then several hundred have been detected by a suite of radio dishes across the Earth, e.g., Parkes, Arecibo, ASKAP, FAST, and CHIME \citep{petroff22AARv}. FRBs are distributed nearly isotropically across the sky and show very large dispersion measures, indicating an extragalactic origin. Hence, their large fluences translate into very bright luminosities, $\sim$9 orders of magnitude brighter than the Crab's giant pulses. While most FRBs appear as single events, a few have associated with a single position on the sky, i.e., repeating FRBs \citep{chime20ApJRFRBs}. 
The origin of FRBs is currently a matter of intense debate, and while many theoretical possibilities exist \citep{platts19PhR}, one of the leading models is a NS or magnetar central engine. 

In a breakthrough discovery, observational evidence for the magnetar model as a source of FRBs occurred on 2020 April 28, when an FRB-like radio burst was detected from the Galactic magnetar SGR 1935+2154 \citep{chime2020:1935, Bochenek20:1935}, in the winding hours of a major burst storm \citep{younes20ApJ1935}; it had a fluence rivaling those of the faint end of extragalactic FRBs. 
Moreover, the FRB occurred simultaneously with a bright, short X-ray burst, connecting it to magnetar activity and providing crucial evidence for its triggering mechanism \citep[e.g.,][]{mereghetti20ApJ,li21NatAs,ridnaia2021NatAs}. In the following years, SGR 1935+2154 has shown several more radio bursts \citep{kirsten2020}, most notably at times of bursting activity. Though most radio bursts occur simultaneous to X-ray bursts, the majority of X-ray bursts occur without a simultaneous radio signal \citep{Lin2020Natur,bailes2021MNRAS} suggesting special circumstances for the emission of FRB-like bursts from magnetars. Indeed, a comparison of the X-ray burst associated with the FRB and NICER+Fermi bursts belonging to the same burst storm of April 2020 reveal the former to have a distinctive spectrum.  This is a clue to either its emission mechanism or triggering locale \citep{younes2021NatAs}, and has been seen in data from INTEGRAL \citep{mereghetti20ApJ}, Fermi/GBM \citep{Lin2020ApJ}, Konus \citep{ridnaia2021NatAs}, and HXMT \citep{li21NatAs}, among others.

\begin{figure*}[t!]
\begin{center}
\includegraphics[width=\linewidth]{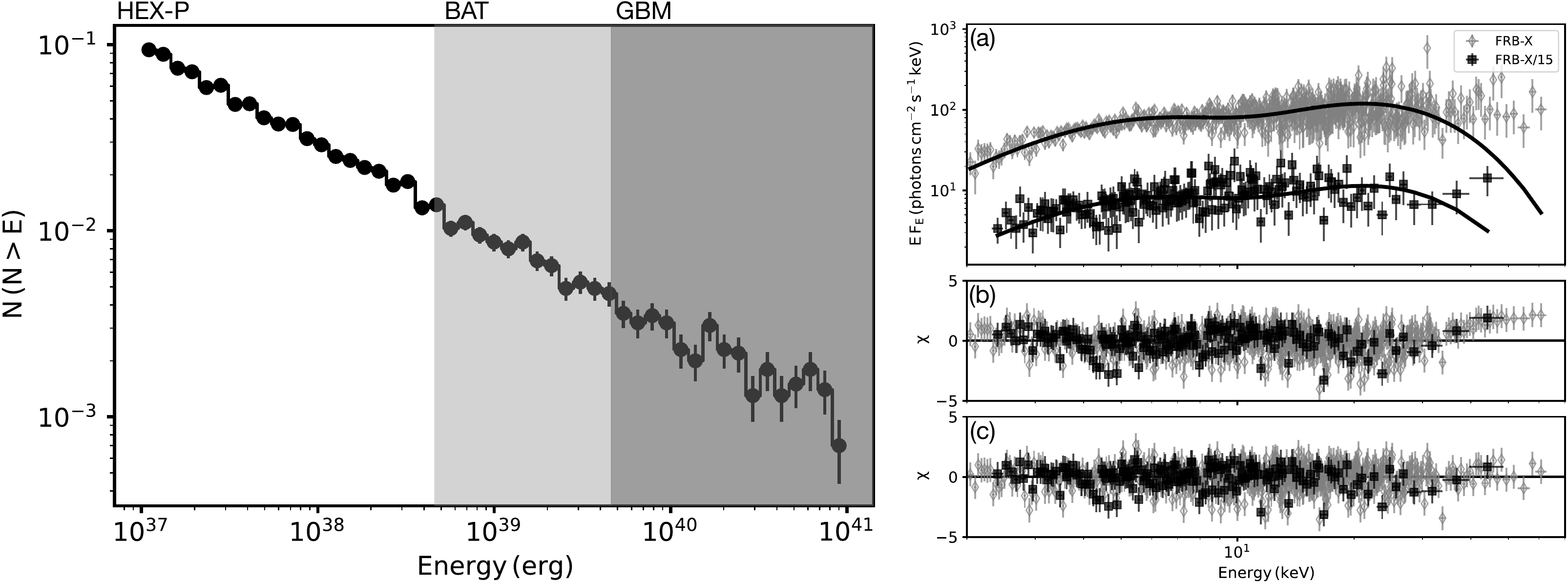}
\end{center}
\caption{{\sl Left panel.} $\log N-\log E$ distribution of magnetar short bursts, where $E$ is the burst total energy. The light and dark shaded areas are the sensitivity limits ($>5\sigma$ detection) of Swift/BAT and Fermi/GBM, respectively, assuming a source distance of 20~kpc. HEX-P is sensitive to the full burst energy distribution for every magnetar in the Galaxy. {\sl Right panel.} Panel (a) shows a HEX-P Xspec simulation of the FRB200428-associated X-ray burst spectrum (gray diamonds) assuming a non-thermal cutoff power-law as measured with HXMT \citep{li21NatAs}. The solid line is the best-fit thermal double blackbody (2BB) model. Black squares are a simulated spectrum with the same assumptions, but with a fluence that is a factor 15 smaller. Panel (b) shows the residuals from the thermal fit in units of 1$\sigma$, with the same grey and black symbols. Panel (c) shows the residuals from the non-thermal fit in units of 1$\sigma$, again with the same grey and black symbols. HEX-P will distinguish the thermal vs non-thermal nature of short X-ray bursts for fluences that are over an order of magnitude weaker than FRB-X (see text).}\label{burstlogNlogS}
\end{figure*}

The discovery of FRB-like bursts from magnetars opened up new avenues for the study of extragalactic FRBs \citep[e.g.,][]{Wadiasingh19ApJ}, radio emission from young INSs \citep{Philippov2022ARAA}, and, more generally, plasma and pair production in magnetar magnetospheres \citep[e.g.][]{mahlmann22ApJ,beloborodov20ApJ,Yuan2020ApJ}. Yet, as is the case with every new fundamental discovery, more questions arise in its aftermath, e.g., (1) What is unique about the FRB-associated X-ray bursts, and why do the majority of X-ray bursts lack a radio counterpart? (2) What is the distribution of the spectral properties for FRB-associated X-ray bursts? Are their distinctive spectral properties universal across radio fluence? (3) Is the radio to X-ray flux ratio ($L_{\rm R}/L_{\rm X}$) constant for all FRB-like radio bursts? (4) What is the radio-X-ray time-lag across burst fluence?

The answers to these open questions are critical for improving our understanding of the FRB phenomenon, both Galactic and extragalactic. Answering these questions will require (1) a broad X-ray coverage given that the spectral energy distributions of magnetar bursts peak in the 20-30 keV range, (2) high timing resolution ($\lesssim1$~ms) for accurately measuring the radio-X-ray lag in burst arrival time \citep{mereghetti20ApJ}, (3) sensitivity to faint X-ray bursts, i.e., fluence $<10^{-7}$~erg~cm$^{-2}$, to sample a large fraction of the X-ray and radio burst fluence distribution given their steep shapes ($N\propto S^{-0.6}$, Figure~\ref{burstlogNlogS}, \citealt{younes20ApJ1935}). HEX-P is the only facility to satisfy all the above criteria. Its only limitation is the small field-of-view, yet, most radio FRB bursts detected from the Galactic magnetar SGR 1935+2154 occurred at the time of major burst storms, which last up to a few days. This is sufficient time for HEX-P to slew to the target. We also note that NuSTAR was observing SGR 1935+2154 at the time of its 2022 October FRB-like burst, though the source was Earth-occulted \citep{dong2022ATel15681,enoto22ATel15690}. Due to the  L1 orbit of HEX-P, such misfortune is naturally avoided. 

The left panel of Figure~\ref{burstlogNlogS} shows a simulated $\log N-\log S$ magnetar burst energy distribution, which follows a power-law of the form $dN/dE\propto E^{-1.6}$ \citep{gogus99ApJ,gogus00ApJ,gavriil04ApJ,horst2012ApJ,younes20ApJ1935}. The Fermi/GBM $>5\sigma$ sensitivity to typical short bursts from a magnetar at a distance of 20~kpc is shown in dark gray \citep{meegan09ApJ} while that for Swift/BAT is shown in light gray \citep{lien16ApJ}. At that distance, HEX-P will detect bursts with energies comparable to the persistent emission, i.e., $\sim10^{37}$~erg (in a 1-second interval), covering significantly more of the short burst energy distribution. This will ensure the detection of X-ray bursts associated with faint radio bursts and provide answers to questions (3) and (4).

To address questions (1) and (2), we performed HEX-P HET simulations of an X-ray burst with spectral properties similar to that of the FRB-associated X-ray burst (which we call FRB-X) as determined by HXMT \citep{li21NatAs}; i.e., a cutoff power-law with $\Gamma=1.6$ and $E_{\rm cut}=80$~keV, and a fluence of $5.0\times10^{-7}$~erg~cm$^{-2}$ (Figure~\ref{burstlogNlogS}, gray diamonds in panel (a)). We then fit this spectrum with a thermal 2 blackbody (2BB) model (shown as a solid black line). The thermal model fails to provide a statistically acceptable fit to the data (panel (b)), unlike the non-thermal model (panel (c)). This is confirmed through Xspec simulations which show that $\gg 99\%$ of simulated spectra drawn from the thermal model have better fit statistics. This indicates that HET alone will discern the non-thermal nature of any bursts similar to FRB-X. We then performed a set of simulations assuming the same spectral model as FRB-X, while decreasing the fluence by increments of factor 2. We then fit each spectrum to a thermal 2BB model, and assess the fit quality through simulations. We find that we can discern (at the $\approx3\sigma$ level) between the thermal and non-thermal model down to a fluence of $\sim 3\times10^{-8}$~erg~cm$^{-2}$, i.e., a factor 15 fainter than FRB-X (Figure~\ref{burstlogNlogS}, black squares).

\section{Central Compact Objects} 

The CCO class of NSs are defined by the following  observational characteristics: steady, soft thermal X-ray emission, lack of a surrounding pulsar wind nebula, and non-detection at all other wavelengths.
X-ray pulses have been detected from only three of the dozen known CCOs \citep{Gotthelf2013}.
Two of these three, the CCOs in the Puppis A and Kes 79 SNRs, have the lowest spin down measured magnetic fields among young neutron stars, with $B_s \sim 3 \times 10^{10}$~G.

While only a dozen CCOs are currently known, their locations in young SNRs indicates they may represent a significant fraction of all NS births.
Hence, understanding how young NSs are born with such small dipole magnetic fields is important to address how CCOs fit within the broad INS family.
Increasing the number of CCO spin period and period derivative measurements is critical. Because CCOs are only detected at X-ray wavelengths, these spin period searches can only be done in the X-ray band. This is an area where HEX-P can shine as a follow-up observatory, capable of both searching for X-ray pulsations and, after identifying a spin period, making the required phase resolved spectroscopic measurements.

\begin{figure}[t!]
\begin{center}
\includegraphics[width=0.5\linewidth]{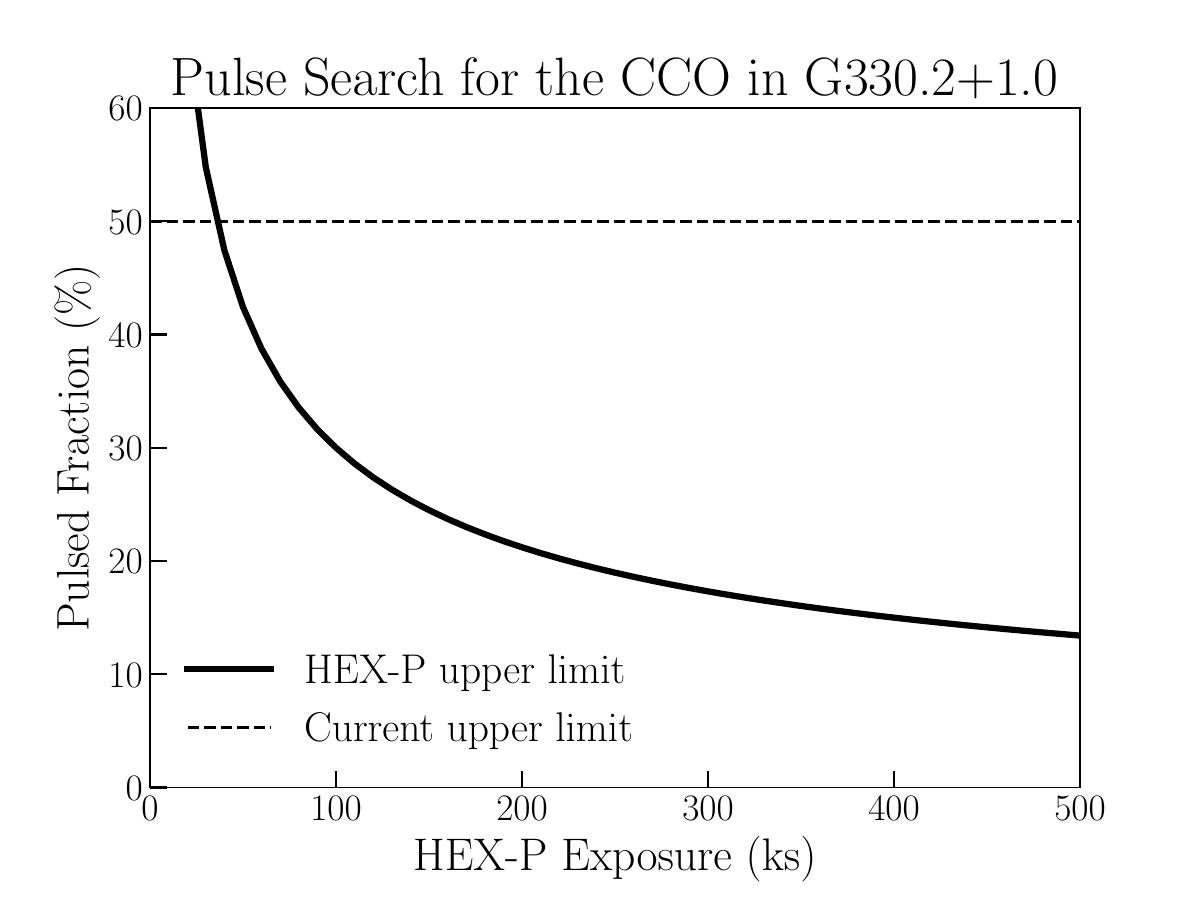}
\end{center}
\caption{HEX-P pulsed fraction upper limits as a function of exposure for the CCO in G330.2+1.0.}\label{fig:CCO}
\end{figure}

As an example, the CCO in G330.2$+$1.0 is a promising target for HEX-P thermal pulse searches. Previous searches with XMM-Newton were limited by the high background from the surrounding SNR thermal emission \citep{Alford2023}. This is highlighted in Equation \ref{eq:z2_to_pf} which relates the calculated pulsed fraction upper limit $f_{p}^{\rm max}$ to the number of source counts $N_{s}$ and background counts $N_{b}$:
\begin{equation}
f_{p}^{\rm max} = 2(1 + N_{b}/N_{s}) \sqrt{P_s/N},
\label{eq:z2_to_pf}
\end{equation}
Since CCOs are found in young supernova remnants, in many cases the thermal background emission from the remnant can be significant, hindering the ability to detect the underlying NS pulsations. To overcome these obstacles, both high X-ray timing and angular resolution are required. HEX-P's 3.5 arcsec PSF (for the LET detector) will allow for a significantly reduced background. Detailed comparisons of the performance of HEX-P in pulsar searches compared to XMM-Newton and NuSTAR can be found in Bachetti et al., in prep. (ULXs and extragalactic pulsars) and Mori et al., in prep. (the Galactic Center).
Figure \ref{fig:CCO} presents the pulsed fraction upper limits as a function of exposure time for the CCO in G330.2$+$1.0.
HEX-P will significantly reduce the current pulsed fraction upper limit, likely leading to a secure determination of the spin period.

Once an X-ray pulse period is found, a measurement of the period derivative can easily follow, as well as the characterization of the timing properties of these sources, e.g., $\dot{E}$, $B$-field strength, and spin down luminosity $\dot{E}$. Moreover, such observations will allow us to study the thermal pulse profiles of new CCOs. Energy-dependent pulse profile modeling is a powerful tool to map the surface thermal emission \citep{Bogdanov2014,Alford2022}. The HEX-P LET has the required effective area, timing resolution, low background, and angular resolution to produce more detailed maps, while increasing the pool of studied sources.

\section{Rotation-Powered Pulsars (RPPs)}

In contrast to the magnetically-powered magnetars and passively cooling CCOs, many pulsars are powered by the loss of their rotational kinetic energy. 
The 33~ms Crab pulsar is perhaps the best known pulsar in this class, with significant rotationally powered emission extending into the hard X-ray band \citep{Madsen2015}. 
HEX-P observations of the Crab pulsar will be a significant improvement over NuSTAR. 
For instance, HEX-P will enable low background phase-resolved studies of the Crab pulsar, by resolving the Crab pulsar from the bright pulsar wind nebula (PWN) background emission.

HEX-P will allow us to observe in detail the faint, middle-aged ($\sim 10^{5}$ yr) pulsars which offer an opportunity to study how pulsars evolve and eventually ``die'', ceasing as X-ray and radio emitting sources. As RPPs age, their X-ray luminosity will decrease with their spin down power, making observations more challenging compared to younger RPPs.

There are three well-known nearby middle-aged pulsars that have been dubbed the ``three musketeers'': PSR B0656$+$14, PSR B1055$-$52 and Geminga. These three pulsars all exhibit thermal surface emission and non-thermal magnetospheric emission. They have similar $\sim 10^{12}$~G spin down magnetic fields and $10^{34}$ erg~s$^{-1}$ spin down luminosities \citep{DeLuca2005}. Despite their relative proximity, open questions remain regarding the physics of their surface thermal emission, and the extent of their non-thermal emission.

\begin{figure}[t!]
\begin{center}
\includegraphics[width=0.7\linewidth]{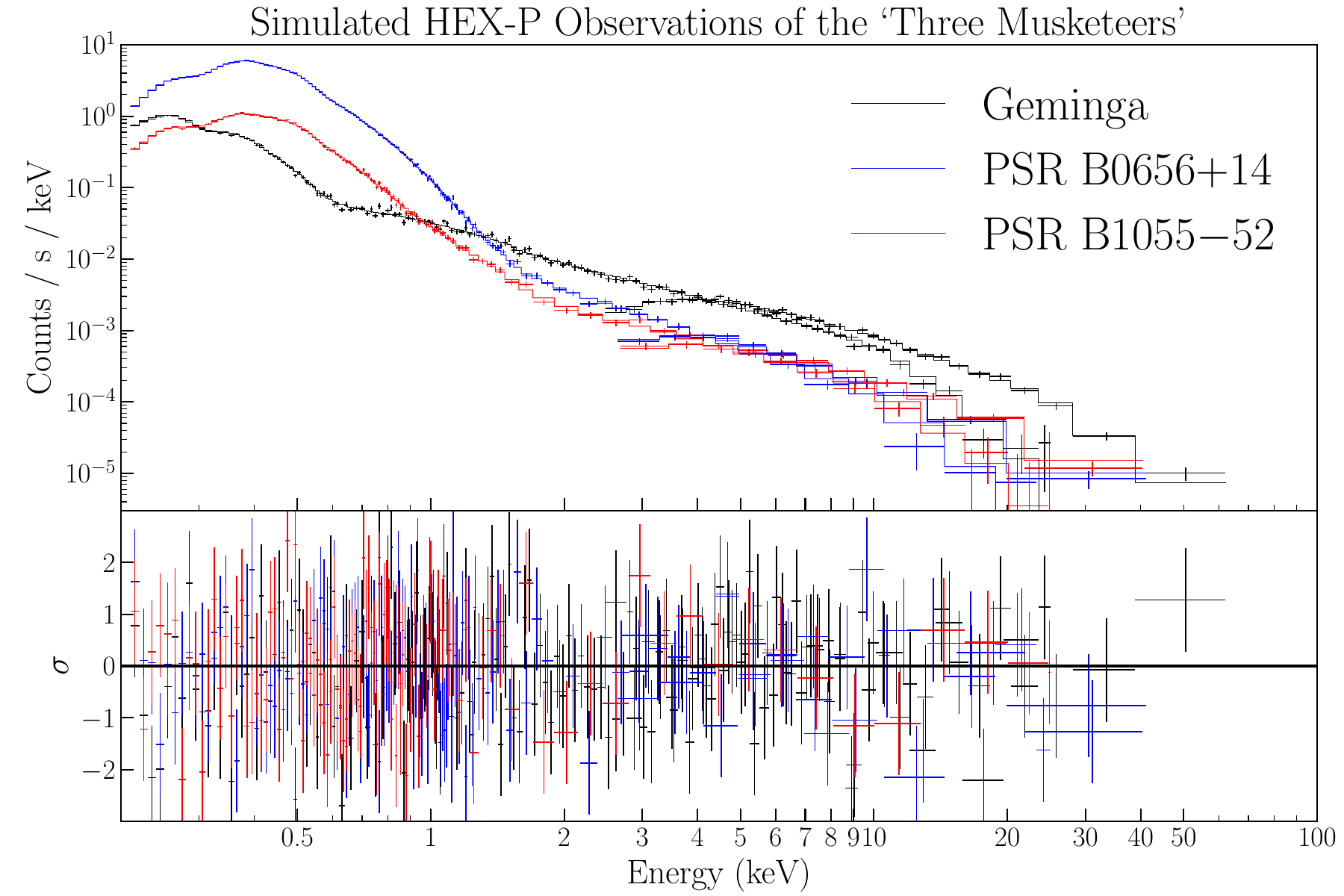} 
\end{center}
\caption{Simulated HEX-P observations of the Geminga pulsar, PSR B0656+14, and PSR B1055-52. 
HEX-P will be capable of extending the current hard X-ray detection limit for Geminga from 20 keV to 50 keV.}\label{fig:geminga}
\end{figure}

Open questions about Geminga are particularly interesting given its potentially large contribution to the local leptonic cosmic ray flux, and its status as the second brightest gamma-ray source in the sky.
\cite{Mori2014} reported on a 150 ks NuSTAR observation of Geminga and found several spectral models were consistent with the data.
Figure \ref{fig:geminga} shows a simulated 200 ks HEX-P observation of Geminga based on the two blackbody plus powerlaw model, with $\Gamma = 1.7$, $kT_{1} =  44 $ eV, and $kT_{2} = 195$ eV.
The cooler thermal component corresponds to emission from the whole NS surface, and the hot component, if its existence is confirmed, may correspond to emission from a hot polar cap.
We find that, if the powerlaw extends to higher energies, HEX-P will extend the detection of non-thermal emission from $\sim$20 keV to $\sim$50 keV.
The X-ray emission of Geminga above 20 keV is unexplored territory, and the detection of changes in the spectrum could be important clues to the physics of its magnetosphere.

Figure \ref{fig:geminga} also shows simulated HEX-P observations of PSR B0656$+$14 and PSR B1055$-$52, performed using the spectral parameters from \cite{DeLuca2005}. We find that 150 ks HEX-P observations will measure the photon indices $\Gamma$ of all three of these faint pulsars to better than $10\%$.

HEX-P will also potentially address a fundamental mystery regarding PSR B0656$+$14, PSR B1055$-$52 and Geminga. B0656$+$14, and PSR B1055$-$52 clearly have small surface thermal hot spots, presumably corresponding to the heated pulsar polar caps. If Geminga has a similar hot spot, then its luminosity is at least two orders of magnitude dimmer \citep{Jackson2005}. 
HEX-P's high throughput and broad coverage of Geminga's X-ray spectrum will allow us to answer this question.

\section{Potential discovery space}

Present and future large field-of-view facilities at all wavelengths from radio to PeV energies will result in a large number of unidentified sources, especially within the Galactic plane. This is already evident at GeV energies and beyond, i.e., Fermi/LAT, H.E.S.S, and LHASSO, where the number of unknown sources outweigh the number of identified ones, a problem that will only be exacerbated with the Cherenkov Telescope Array Observatory (CTAO). These high energy sources represent the most efficient particle accelerators in the universe, and for the Galactic ones, their most likely counterpart involves a pulsar (Mori et al. 2023 (in prep.)). At X-ray energies, eROSITA (\citealt{Predehl2021AA}, and potentially STAR-X~\citealt{2022HEAD...1910845Z}) will provide some of the deepest wide-field X-ray surveys of our Galaxy, with a $>10 \times$ increase in the number of X-ray sources compared to ROSAT. A large fraction of these X-ray sources will be of unknown origin, and a non-negligible fraction should be INSs \citep{Pires2017AN}. Deeper, targeted exposures, as possible with HEX-P, will be required to identify them. Finally, in the radio, the Square Kilometer Array and Deep Synoptic Array 2000 are expected to increase the number of currently known pulsars by a factor of 10 (i.e., to $\sim$20,000 pulsars), as well as detect a large number of new SNR shells and candidate wind nebulae.

Simply detecting X-ray point sources in the error region of unidentified gamma-ray and radio sources will not yield a secure identification. Furthermore, providing a high quality soft X-ray spectrum of eROSITA/STAR-X sources will not be enough to firmly distinguish their origin. High timing resolution is required to discern the pulsar nature from other types of X-ray emitters, such as low-mass X-ray binaries, cataclysmic variables, background active galaxies, etc. Moreover, being limited to a soft X-ray detector will hamper our ability to probe deep into the Galaxy due to absorption, stressing the need for hard X-ray coverage. Last but not least, high spatial resolution is required for crowded regions such as Sgr A$^\star$ (Mori et al. 2023b (in prep.)), and to isolate the pulsar emission from any surrounding SNR and/or nebula (as noted in \S5 with regards to candidate CCOs). These issues have already been demonstrated with NuSTAR. For instance, although a candidate soft X-ray counterpart to the TeV $\gamma$-ray source HESS J1640-465 had been proposed, only NuSTAR was able to detect pulsed emission confirming its pulsar nature \citep{Gotthelf2019}. This was mainly due to the combination of heavy absorption ($10^{23}$~cm$^{-2}$) in the source direction and non-negligible contamination by the PWN \citep{Gotthelf2019}. Continued NuSTAR follow-up of this pulsar found the braking index $n>3$, possibly pointing to a magnetic quadrupole in the source \citep{Archibald2016ApJ}.

HEX-P presents the ideal satellite to follow-up Galactic gamma-ray sources, INS candidates from wide-area X-ray surveys, and pulsar/pulsar candidates from deep radio surveys. 
Increasing the number of isolated X-ray pulsars, such as XDINs, CCOs, RPPs, and magnetars, will enable a deeper understanding of their physics, X-ray properties, environment (e.g.,  wind nebulae, SNRs and nearby cosmic ray acceleration sites), and progenitors, in turn teaching us about NS formation and evolutionary tracks through, e.g., population synthesis modeling \citep[e.g.,][]{Gullon2015MNRAS,Dirson2022AA} and magneto-thermal evolution models \citep[e.g.,][]{vigano13MNRAS,gourgouliatos2016PNAS}.

\newpage

\section*{Conflict of Interest Statement}

The authors declare that the research was conducted in the absence of any commercial or financial relationships that could be construed as a potential conflict of interest.

\section*{Author Contributions}

This paper is the result of the work of the HEX-
P pulsar/magnetar working group, led by JA and GY.
JA and GY designed and performed simulated HEX-P observations.
JA, GY, and ZW wrote the manuscript.
All authors have provided useful advice during working group
discussions and comments on the manuscript.

\section*{Acknowledgments}
 The work of D.~S. was carried out at the Jet Propulsion Laboratory, California Institute of Technology, under a contract with NASA. Z.~W. acknowledges support by NASA under award number 80GSFC21M0002. This work has made use of the NASA Astrophysics Data System.

\bibliographystyle{Frontiers-Harvard}
\bibliography{test}
\end{document}